\documentstyle [12pt] {article}
\input epsf
\newcommand{\gtwid}{\mathrel{\raise.3ex\hbox{$>$\kern-.75em\lower1ex
\hbox{$\sim$}}}}
\newcommand{\ltwid}{\mathrel{\raise.3ex\hbox{$<$\kern-.75em\lower1ex
\hbox{$\sim$}}}}
\newcommand{\beq}{\begin{equation}}
\newcommand{\eeq}{\end{equation}}
\newcommand{\beqs}{\begin{eqnarray}}
\newcommand{\eeqs}{\end{eqnarray}}

\begin{document}

\def\thefootnote{\fnsymbol{footnote}}
\baselineskip 6.0mm

\begin{flushright}
ITP-SB-97-11 \\ 
March, 1997
\end{flushright}

\vspace{4mm}
\begin{center}

{\Large \bf Families of Graphs With }

\vspace{2mm}

{\Large \bf Chromatic Zeros Lying on Circles} 

\vspace{8mm}

\setcounter{footnote}{0}
Robert Shrock\footnote{email: shrock@insti.physics.sunysb.edu}
\setcounter{footnote}{6}
and Shan-Ho Tsai\footnote{email: tsai@insti.physics.sunysb.edu}

\vspace{6mm}

Institute for Theoretical Physics  \\
State University of New York       \\
Stony Brook, N. Y. 11794-3840  \\

\vspace{10mm}

{\bf Abstract}
\end{center}
   We define an infinite set of families of graphs, which we call $p$-wheels 
and denote $(Wh)^{(p)}_n$, that generalize the wheel ($p=1$) and biwheel 
($p=2$) graphs.  The chromatic polynomial for $(Wh)^{(p)}_n$ is calculated, 
and remarkably simple properties of the chromatic zeros are found: (i) the 
real zeros occur at $q=0,1,...p+1$ for $n-p$ even and $q=0,1,...p+2$ for 
$n-p$ odd; and (ii) the complex zeros all lie, equally spaced, on the unit 
circle $|q-(p+1)|=1$ in the complex $q$ plane.  In the $n \to \infty$ limit, 
the zeros on this circle merge to form a boundary curve separating two regions
where the limiting function $W(\{(Wh)^{(p)}\},q)$ is analytic, viz., the 
exterior and interior of the above circle.  Connections with statistical
mechanics are noted. 

\vspace{16mm}

\pagestyle{empty}
\newpage

\pagestyle{plain}
\pagenumbering{arabic}
\renewcommand{\thefootnote}{\arabic{footnote}}
\setcounter{footnote}{0}

\section{Introduction}

   The chromatic polynomial $P(G,q)$ of a graph $G$ plays a fundamental role in
graph theory; it gives the number of ways of coloring this graph $G$ with $q$
colors such that no two adjacent vertices have the same color
\cite{birk}-\cite{graphs}.  Since $P(G,q) \le q^n$ for a graph $G$ with $n$ 
vertices, it is natural to consider the asymptotic limiting function 
\beq
W(\{G\},q) = \lim_{n \to \infty} P(G,q)^{1/n}
\label{w}
\eeq
where $\{G\}$ denotes the limit as $n \to \infty$ of the family of 
$n$-vertex graphs of type $G$.  There is a natural generalization, which we 
assume here, of the variable $q$ from integer to complex values.  Recently, 
we calculated the $W(\{G\},q)$ functions for several families of graphs 
$\{G\}$ \cite{w} and studied their analytic properties.  One of the 
interesting results that we found was that for three families of graphs,
namely, circuit, wheel, and biwheel graphs the zeros of the respective 
chromatic polynomials, i.e., the chromatic zeros of these graphs, 
lie at certain real integer values and on certain unit circles in the 
complex $q$ plane.  A question that occurred to us was the following: 
Can one generalize this finding and if so, can one thereby
achieve a deeper understanding of the result.  We have succeeded in finding a 
comprehensive generalization which does yield further insight into this result,
and we present this here.  Given the elementary relation that the chromatic
polynomial $P(G,q) = Z(G,q,T=0)_{PAF}$, where $Z(G,q,T=0)_{PAF}$ is the
partition function of the zero-temperature $q$-state Potts antiferromagnet (AF)
on the graph $G$, and the consequent relation, for the thermodynamic limit, 
between $W(\{G\},q)$ and the exponent of the ground state entropy of the Potts
AF, our results also have implications for statistical mechanical properties 
of the Potts antiferromagnet on a certain family of graphs. 

   To construct the $p$-wheel with $n$ vertices, we start with a circuit with 
$n-p$ vertices, $C_{n-p}$, and adjoin $p$ points to each of the points on the 
circuit and to each other.  Here, obviously, $n-p \ge 1$ and, for a nontrivial
circuit, $n-p \ge 2$.  This process could
be symbolized as the product $K_p \times C_{n-p}$, where $K_p$ is the complete
graph on $p$ vertices.  We denote the $p$-wheel as $(Wh)^{(p)}_n$.  This
generalizes the definitions of the $n$-vertex wheel and biwheel graphs $(Wh)_n$
and $U_n$; these are seen now as special cases of the $p$-wheel for the 
respective values $p=1$ and $p=2$: $(Wh)^{(1)}_n=(Wh)_n$ and 
$(Wh)^{(2)}_n=U_n$.  The $n$-vertex circuit graph $C_n$ is subsumed as the 
$p=0$ special case: $(Wh)^{(0)}_n = C_n$. 
Thus, each family $K_p \times C_{n-p} = 
(Wh)^{(p)}_n$ forms an infinite set as one increases the number of vertices
$n$.  
Clearly, 
\beq
(Wh)^{(p)}_{p+1} = K_{p+1}
\label{wkrel}
\eeq

\section{Theorems on Chromatic Zeros of $(Wh)^{(p)}_n$ }

   In general, the chromatic polynomial $P(G,q)$ for a graph $G$ has zeros at
the discrete values $q = 0$, $q=1$, and, if $G$ contains one or more triangles,
also the value $q=2$.  Furthermore, since $P(G,q)$ is real for real $q$, it 
follows that 
the zeros of this polynomial (i.e. the chromatic zeros) are invariant under
the complex conjugation $q \to q^*$.  We consider an $n$-vertex graph $G$ in a 
particular family $\{ G \}$ of graphs for which, as $n \to \infty$, aside 
from the above-mentioned discrete zeros at $q=0,1$ and, for $G \supseteq
\bigtriangleup$, $q=2$, the remaining zeros merge to form the union of 
boundaries ${\cal B}$ separating various regions in the complex $q$
plane. \cite{spec} 
A general question that one may ask about this latter set of zeros is 
whether all, or a subset of them, lie exactly on the boundary curves 
${\cal B}$ even for finite $n$. (Some of the set of discrete zeros may also 
lie on ${\cal B}$.)  Clearly, the fact that a subset of zeros merge to form 
the boundary ${\cal B}$ in the $n \to \infty$ limit
does not imply that, for finite graphs $G$, some subset of zeros will lie 
precisely on ${\cal B}$.  We have investigated this question and have found
that there do exist some families of graphs $\{G\}$ for which the zeros of 
$P(G,q)$ (aside from a certain discrete set) lie exactly on the respective 
boundary curves ${\cal B}$.  Furthermore, we find that these families of graphs
also provide an answer to the question posed at the beginning of this paper,
concerning a generalization of the property that ${\cal B}$ comprises a (unit) 
circle.  

   Our main results are contained in the following two theorems 
(as noted above, for a nontrivial circuit $C_{n-p}$ in $(Wh)^{(p)}_n$, 
$n-p \ge 2$). 

\begin{flushleft}
Theorem 1
\end{flushleft}
The chromatic polynomial for the $p$-wheel is
\beq
P((Wh)^{(p)}_n,q) = \Bigl [ \ \prod_{\ell=0}^{p-1}(q-\ell) \ \Bigr ]
\Bigl [ \ (q-p-1)^{n-p} + (-1)^{n-p}(q-p-1) \Bigr ]
\label{pwhp}
\eeq

\vspace{2mm}
\begin{flushleft}
Proof
\end{flushleft}
We use the theorem that if a graph $H$ is obtained from a graph $G$ by 
adding a vertex which is adjacent to all of the vertices of $G$ (i.e. 
connected to them by bonds of the graph), then (e.g., Ref. \cite{rtrev})
\beq
P(H,q) = qP(G,q-1)
\label{adjoin}
\eeq
We take $G=C_{n-p}$, the circuit with $n-p$ vertices, and apply (\ref{adjoin}) 
iteratively $p$ times to get
\beq
P((Wh)^{(p)}_n,q) = 
\Bigl [ \ \prod_{\ell=0}^{p-1}(q-\ell) \ \Bigr ] P(C_{n-p}, q-p) 
\label{pwhpa}
\eeq
Substituting the basic result $P(C_n,q) = (q-1)^n + (-1)^n (q-1)$ then yields
(\ref{pwhp}).  $\Box$. 

Next, we have 

\begin{flushleft}
Theorem 2 
\end{flushleft}
The real zeros of the chromatic polynomial of the $p$-wheel,
$P((Wh)^{(p)}_n,q)$ occur at $q=0,1,..., p+1$
for $n-p$ even and $q=0,1,...,p+2$ for $n-p$ odd, while the complex zeros
all lie on the unit circle $|q-(p+1)|=1$ in the complex $q$ plane. 
Furthermore, the zeros are equally spaced around this circle,
and in the $n \to \infty$ limit, the density $g(\{(Wh)^{(p)}\},\theta)$ of
zeros on the circle $q = p+1 + e^{i\theta}$, $-\pi < \theta \le
\pi$, is a constant, independent of $\theta$.

\vspace{2mm}

\begin{flushleft}
Proof
\end{flushleft}
Defining $x=q-(p+1)$, we can write the second expression in eq. 
(\ref{pwhpa}) as 
\beq
x(x^{n-p-1}+(-1)^{n-p})
\label{xterm}
\eeq
It follows that the complex zeros of $P((Wh)^{(p)}_n,q)$ are the complex
solutions to the equation $x^{n-p-1}+(-1)^{n-p}=0$, which clearly lie on the
unit circle $|x|=1$, i.e., $|q-(p+1)|=1$.  Specifically, for $n-p$ even,
$x=-1$, i.e., $q=p$ is a root, and the $n-p-2$ complex roots are given by
\beq
q = p+1 + e^{i \pi + \frac{2s \pi i}{n-(p+1)}} \ , \qquad s = 1,.., n-p-2
\label{qsolnpeven}
\eeq
while for $n-p$ odd, $x=\pm 1$, i.e., $q=p, \ p+2$ are real roots, while the
$n-p-3$ complex roots are
\beq
q = p+1 + e^{\frac{2s \pi i}{n-(p+1)}} \ , \qquad s = 1,..,n-p-2 \ ; s \ne
\frac{n-p-1}{2} 
\label{qsolnpodd}
\eeq
where the (integer) value $\frac{(n-p-1)}{2}$ is omitted because this root is 
real.  These
complex roots obviously come in complex conjugate pairs; a factorization which
makes this explicit is, for $n-p$ even, 
\beq
P((Wh)^{(p)}_n,q)_{n-p \ {\rm even}} = 
\biggl [ \ \prod_{\ell=0}^{p+1}(q-\ell) \ \biggr ]
\prod_{j=0}^{\frac{n-p}{2} - 2}
\Bigl \{ q-(p+1+e^\frac{(2j+1)\pi i}{n-(p+1)}) \Bigr \}
\Bigl \{ q-(p+1+e^\frac{-(2j+1)\pi i}{n-(p+1)}) \Bigr \}
\label{pneven}
\eeq
and for $n-p$ odd, 
\beq
P((Wh)^{(p)}_n,q)_{n-p \ {\rm odd}} 
= \biggl [ \ \prod_{\ell=0}^{p+2}(q-\ell) \ \biggr ]
\prod_{j=1}^{\frac{n-p-1}{2}-1}
\Bigl \{ q-(p+1+e^\frac{2j\pi i}{n-(p+1)}) \Bigr \}
\Bigl \{ q-(p+1+e^\frac{-2j\pi i}{n-(p+1)}) \Bigr \}
\label{pnodd}
\eeq
The complex factors in eqs. (\ref{pneven}) and (\ref{pnodd}) follow 
immediately from eqs. (\ref{qsolnpeven}) and (\ref{qsolnpodd}).   
This proves the second part of our theorem,
that the complex roots lie on the unit circle $|x|=1$ with equal spacing.  In
particular, in the $n \to \infty$ limit, it follows that the density
$g(\{(Wh)^{(p)}\},\theta)$ of zeros on the circle $q=p+1+e^{i\theta}$, $-\pi <
\theta \le \pi$ is a constant, independent of $\theta$.  If one normalizes this
density according to 
\beq
\int_{-\pi}^{\pi} g(\{(Wh)^{(p)}\},\theta) \ d\theta = 1
\label{gnorm}
\eeq
then
\beq
g(\{(Wh)^{(p)}\},\theta) = \frac{1}{2\pi} 
\label{gg}
\eeq
Concerning the product of real integer roots, we
observe that for $n-p$ even, eq. (\ref{xterm}) has the real factors $x$ and
$x+1$, i.e., $q-(p+1)$ and $q-p$ which, combined with the factor
$\prod_{\ell=0}^{p-1}(q-\ell)$ in (\ref{pwhp}), yield the factor 
$\prod_{\ell=0}^{p+1}(q-\ell)$ in (\ref{pneven}).  For $n-p$ odd,
eq. (\ref{xterm}) has the real factors $x$, $x+1$, and $x-1$, i.e., $q-(p+1)$,
$q-p$, and $q-(p+2)$ which, combined again with the factor 
$\prod_{\ell=0}^{p-1}(q-\ell)$ in (\ref{pwhp}), yield the factor 
$\prod_{\ell=0}^{p+2}(q-\ell)$ in (\ref{pnodd}).  This proves the first part of
the theorem, which specifies the real roots of $P((Wh)^{(p)}_n,q)$.  This
completes the proof of both parts of the theorem.  \ \ $\Box$

\section{Analytic Structure of $W(\{(Wh)^{(p)}\},q)$}

   We next consider the limiting function $W(\{(Wh)^{(p)}\},q)$.  
As we have discussed in Ref. \cite{w}, eq. (\ref{w}) is not, in general, 
adequate to define this function because of a noncommutativity that occurs at
certain special values of $q$, which we shall denote $q_s$, where
\beq
\lim_{n \to \infty} \lim_{q \to q_s} P(G,q)^{1/n} \ne
\lim_{q \to q_s} \lim_{n \to \infty} P(G,q)^{1/n}
\label{wnoncomm}
\eeq
The origin of this noncommutativity of limits
is an abrupt change in the behavior of $P(G,q)$ in the vicinity of such a
point $q_s$; for $q \ne q_s$, $P(G,q)$ grows exponentially as the number of
vertices $n$ in $G$ goes to infinity: $P(G,q) \sim a^n$ for some nonzero $a$,
whereas precisely at $q=q_s$, it has a completely different type of behavior; 
typically, $P(G,q_s) = c_0(q_s)$ where $c_0(q)$ may be a constant, independent
of $n$ or may depend on $n$, but in a way that does not involve exponential 
growth, like $(-1)^n$. The set of special points $\{q_s\}$ includes $q=0$, 
$q=1$, and, if $G \supseteq \bigtriangleup$, also $q=2$; at these points, 
$P(G,q_s)=0$.

   In the present case, we find that for a fixed $p$, in the $n \to \infty$
limit, the zeros on the unit circle $|q-(p+1)|=1$ merge to form a (complete) 
circle with a density given by (\ref{gg}).  This circle thus constitutes the
boundary curve ${\cal B}$ and divides the complex $q$ plane into two regions:
\beq
R_1 : \ q \quad {\rm such \ \ that} \quad |q-(p+1)| > 1 \quad 
\label{r1}
\eeq
and 
\beq
R_2 : \ q \quad {\rm such \ \ that} \quad |q-(p+1)| < 1
\label{r2}
\eeq
The set of special points $\{q_s\}$ associated with the noncommutativity in 
(\ref{wnoncomm}) are the real integral zeros of $P((Wh)^{(p)}_n,q)$, i.e., 
$\{q_s\} = 0,1,...,p+1$ for $n-p$ even and $\{q_s\}=0,1,...,p+2$ for $n-p$ 
odd.  We find that
\beq
W(\{(Wh)^{(p)}\},q) = q-(p+1) \qquad {\rm for} \quad q \in R_1 \quad 
{\rm and} \quad q \notin \{q_s\} 
\label{wr1}
\eeq
Because of the noncommutativity (\ref{wnoncomm}), 
the formal definition (\ref{w}) is, in
general, insufficient to define $W(\{G\},q)$ at the set of special points
$\{q_s\}$; at these points, one must also specify the order of the limits in
(\ref{wnoncomm}).  One can maintain the analyticity of $W(\{G\},q)$ at these
special points $q_s$ of $P(G,q)$ by choosing the order of limits in the
right-hand side of eq. (\ref{wnoncomm}):
\beq
W(\{G\},q_s)_{D_{qn}} \equiv \lim_{q \to q_s} \lim_{n \to \infty} P(G,q)^{1/n}
\label{wdefqn}
\eeq
As indicated, we shall denote this definition as $D_{qn}$
Although this definition
maintains the analyticity of $W(\{G\},q)$ at the special points $q_s$, it
produces a function $W(\{G\},q)$ whose values at the points $q_s$ differ
significantly from the values which one would get for $P(G,q_s)^{1/n}$ with
finite-$n$ graphs $G$.  The definition based on the opposite order of limits,
\beq
W(\{G\},q_s)_{D_{nq}} \equiv \lim_{n \to \infty} \lim_{q \to q_s} P(G,q)^{1/n}
\label{wdefnq}
\eeq
gives the expected results like $W(G,q_s)=0$ for $q_s=0,1$,
and, for $G \supseteq \triangle$, $q=2$, but yields a function $W(\{G\},q)$ 
with discontinuities at the set of points $\{q_s\}$.  
In the present case, we have 
\beq
W(\{(Wh)^{(p)}\},q)_{D_{nq}} = 0 \quad {\rm for} \quad 
q \in R_1 \quad {\rm and} \quad q \in \{q_s\}
\label{wr1qsdnq}
\eeq
while
\beq
W(\{(Wh)^{(p)}\},q)_{D_{qn}} = q-(p+1) \quad {\rm for} \quad q \in R_1 \quad 
{\rm and} \quad q \in \{q_s\}
\label{wr1qsdqn}
\eeq
For $q \in R_2$ there is no canonical choice of phase to use in the $n'th$ root
(\ref{w}) \cite{w}; however, for the complex amplitude we have 
\beq
|W(\{(Wh)^{(p)}\},q)| = 1 \quad {\rm for} \quad q \in R_2 \quad {\rm and} \quad
q \notin \{q_s\}
\label{wr2}
\eeq
The region $R_2$ contains the point $q=p+1$ which is a member of the set of
discrete zeros $\{q_s\}$; for $q=p+1$, the two orders of limits give 
$W(\{(Wh)^{(p)}\},q)_{D_{nq}} = 0$ and 
$|W(\{(Wh)^{(p)}\},q)_{D_{qn}}|=1$. 
The limiting function $W(\{(Wh)^{(p)}\},q)$ is nonanalytic on the boundary
circle $|q-(p+1)|=1$. 

    In Ref. \cite{w} (see also Ref. \cite{p3afhc}) we discussed the quantity
$q_c(\{G\})$, the maximal real point of nonanalyticity of $W(\{G\},q)$, and
inferred the values of $q_c$ for certain lattices $\{G\}=\Lambda$.  A 
corollary of our theorems above is that 
\beq
q_c(\{(Wh)^{(p)}\}) = p+2
\label{qcpwheel}
\eeq
As was the case with the lattices previously considered, we infer that for 
integer $q$ above this critical value, i.e, here, $q > p+2$, 
the Potts antiferromagnet on the infinite-$n$ limit of the 
$p$-wheel graph has a nonzero ground state entropy 
$S_0 = k_B \ln W(\{(Wh)^{(p)}\})$.  For integer
$q \le p-1$, the model is frustrated because of the impossibility of assigning
different colors (i.e., spin states in the Potts AF) to all of the adjacent 
vertices of the $K_p$ subgraph.  It would be interesting to investigate further
the properties of the Potts antiferromagnet on the infinite-$n$ limits of 
$p$-wheel graphs.  

   Because of the connection to statistical mechanics, it is worthwhile to 
contrast our theorem with the famous Yang-Lee circle theorem \cite{yl}. Yang
and Lee considered the Ising model partition function 
$Z=\sum_{\{\sigma_r\}} e^{-\beta {\cal H}}$ at temperature $T$ in an external 
magnetic field $H$, for which the Hamiltonian is
${\cal H} = -\sum_{r,r'} \sigma_r J_{r,r'}\sigma_{r'} - H\sum_r \sigma_r$,
where $\sigma_r = \pm 1$ and $\beta = (k_B T)^{-1}$.  They proved that 
for an arbitrary (finite as well as infinite) lattice with spin-spin couplings
which are ferromagnetic, $J_{r,r'} \ge 0$, but not necessarily nearest-neighbor
(and physical temperature $0 \le T \le \infty$), the zeros of $Z$ in the 
variable $\mu=e^{-2\beta H}$ lie on the unit circle $|\mu|=1$ in the 
$\mu$ plane. In the thermodynamic limit, these zeros merge, becoming dense 
and forming an arc $\mu = e^{i\theta}$ for $\theta_0(T) \le |\theta| \le \pi$,
where $\pm \theta_0(T)$ denote the angles at the endpoints of this arc. 
The free energy is nonanalytic
across the arc.  For temperature $T > T_c$, the arc leaves a gap 
which includes the positive real axis in the $\mu$ plane, but as $T$ 
decreases through the critical temperature $T_c$, this arc closes, pinching 
the real axis at $\mu=1$.  For $0 \le T \le T_c$ the zeros continue to form a 
full closed unit circle. Our theorem shares with the Yang-Lee theorem the 
property of zeros lying on a unit circle and of being true for
finite graphs as well as in the limit $n \to \infty$.  However, there are
several important differences: (i) first, in the Yang-Lee circle theorem, all
of the zeros of $Z$ lie on the 
fixed unit circle $|\mu|=1$, whereas in contrast,
in our theorem, only a subset do and, moreover, the center of the circle is at
a different position, $q=p+1$ for each respective family of $p$-wheel graphs,
$(Wh)^{(p)}_n$; (ii) the Yang-Lee theorem describes a multivariable problem,
dealing with the zeros of $Z$ in the complex $\mu$ plane, which depend
parametrically on the temperature $T$, and it requires in its premise
that all $J_{r,r'} \ge 0$, whereas in our case, once $p$ is specified, i.e.,
one specializes to a given $p$-wheel graph $(Wh)^{(p)}_n$, the zeros
in the complex variable, $q$ are fixed; 
(iii) thus, while in the $n \to \infty$ limit, the Yang-Lee zeros
merge to form an arc which does not form a closed circle for $T > T_c$, the
subset of chromatic zeros that we show lie on the unit circle $|q-(p+1)|=1$
always close to form a complete circle in the $n \to \infty$ limit; and (iv)
on the Yang-Lee arc, the density of zeros is not, in general, a constant,
independent of $\theta$, whereas the density $g(\{(Wh)^{(p)}\},\theta)$ in our
case is, in fact, independent of $\theta$.  Despite 
these differences, the properties that our theorem and the Yang-Lee circle
theorem share in common, namely that they both apply to a finite graph and
imply zeros lying on a unit circle, show a very interesting connection
between these results.

\section{Conclusions}
    
   In conclusion, we have constructed a set of families of 
graphs which we call $p$-wheels, $(Wh)^{(p)}_n$, and calculated their
chromatic polynomials.  We have proved that these provide a (countably
infinite) generalization of the behavior that we had found previously for the
circuit, wheel, and biwheel graphs (which are seen as the special cases $p=0,1$
and 2, respectively), namely that a subset of the chromatic zeros
lie exactly on the unit circle $|q-(p+1)|=1$ in the complex $q$ plane, and 
in the limit as the number of vertices $n \to \infty$, these zeros merge
to form the full circle $|q-(p+1)|=1$ with density independent of position on
the circle.  Since the chromatic zeros lie exactly on this circle even for
finite $n$, the families also constitute a constructive answer to another
question that we posed, namely to find families of graphs such that chromatic
zeros lie precisely on the asymptotic boundary curve ${\cal B}$.  We have
compared our results with the Yang-Lee circle theorem and remarked on the
connections with zero-temperature Potts antiferromagnets.  Our present
theorems provide a deeper and unified understanding of our previous findings
in Ref. \cite{w}.  

This research was supported in part by the NSF grant PHY-93-09888.

\vfill
\eject

\vspace{6mm}

\begin{thebibliography}{99}

\bibitem{birk}{G. D. Birkhoff, Ann. of Math. {\bf 14}, 42 (1912).}

\bibitem{whit}{H. Whitney, Ann. of Math. {\bf 33}, 688 (1932); 
Bull. Am. Math. Soc. {\bf 38}, 572 (1932).}

\bibitem{bl}{G. D. Birkhoff and D. C. Lewis, Trans. Am. Math. Soc.
{\bf 60}, 355 (1946).}

\bibitem{rtrev}{R. C. Read and W. T. Tutte, ``Chromatic Polynomials'', 
in {\it Selected Topics in Graph Theory, 3}, eds. L. W. Beineke and 
R. J. Wilson (Academic Press, New York, 1988).}

\bibitem{graphs}{W. T. Tutte {\it Graph Theory}, vol. 21 of 
{\it Encyclopedia of Mathematics and its Applications}, ed. Rota,
G. C. (Addison-Wesley, New York, 1984).} 

\bibitem{w}{R. Shrock and S.-H. Tsai, Phys. Rev. {\bf E55} (1997), in press.}

\bibitem{spec}{This sort of behavior does not occur for all graphs; for
example, the chromatic zeros of the complete graph on $n$ vertices, $K_n$,
occur at the values $0,1,...,n-1$ and hence do not contain a subset which 
merge to form a continuous boundary curve in the $n \to \infty$ limit.} 

\bibitem{p3afhc}{R. Shrock and S.-H. Tsai, J. Phys. A {\bf 30}, 495 (1997).}

\bibitem{yl}{C. N. Yang and T. D. Lee, Phys. Rev. {\bf 87}, 404 (1952); 
T. D. Lee and C. N. Yang, Phys. Rev. {\bf 87}, 410 (1952).}

\end{thebibliography}
\end{document}